# A Large-Scale GaP-on-Diamond Integrated Photonics Platform for NV Center-Based Quantum Information

Michael Gould,[1,*] Srivatsa Chakravarthi,[1] Ian R. Christen,[2] Nicole Thomas,[1] Shabnam Dadgostar,[3] Yuncheng Song,[4] Minjoo Larry Lee,[4] Fariba Hatami,[3] Kai-Mei C. Fu[1,2]

[1]Department of Electrical Engineering, University of Washington, 185 Stevens Way, Seattle WA 98195-2500
[2]Department of Physics, University of Washington, 3910 15[th] Ave NE, Seattle WA 98195-1560
[3]Department of Physics, Humboldt-Universität zu Berlin, Newtonstrasse 15, 12489 Berlin
[4]Department of Electrical Engineering Yale University, 15 Prospect St., New Haven, CT 06511
*Corresponding author: mike.gould23@gmail.com

**Abstract:** We present chip-scale transmission measurements for three key components of a GaP-on-diamond integrated photonics platform: waveguide-coupled disk resonators, directional couplers, and grating couplers. We also present proof-of-principle measurements demonstrating nitrogen-vacancy (NV) center emission coupled into selected devices. The demonstrated device performance, uniformity and yield place the platform in a strong position to realize measurement-based quantum information protocols utilizing the NV center in diamond.

http://dx.doi.org/10.1364/AO.99.099999

## 1. INTRODUCTION

The negatively charged nitrogen-vacancy (NV) center in diamond is a promising candidate for scalable, measurement-based quantum computation (MBQC) [1–3], wherein quantum entanglement of electron spins is used as the computational resource. Recent years have seen several important experimental results laying the foundations for diamond-based MBQC. These include increasing the electron spin coherence time to 0.6 s [4], coupling the electron spin to long-lived nuclear spins [5–7] and the successful implementation of 2-qubit MBQC protocols [8–10].

The largest remaining hurdle in the path to many-qubit entanglement based on NV centers is photon collection efficiency. In fact, the highest reported success rate for spin entanglement generation between two NV centers is 0.004 Hz [9], well below the electron spin de-coherence rate. The success rate of this free-space protocol was limited by the efficiency with which the NV centers emitted useful photons into the optical collection mode. We will call this quantity the total quantum efficiency. The total quantum efficiency in the referenced work was approximately 0.15%, with the success rate scaling as the square of the efficiency.

Integrated photonic devices are likely the only way to overcome the quantum efficiency problem. Here we detail chip-scale fabrication and characterization of three key components of an integrated diamond photonics platform: waveguide-coupled disk resonators, directional couplers, and grating couplers (Figure 1). The measured device performance suggests possible total quantum efficiency as high as 5.5% with off-chip coupling, and as high as 33% into a traveling waveguide mode (on-chip). Demonstrated device yields and cross-chip uniformity should enable integration of more complex optical circuits in the platform. We also present proof-of-principle measurements demonstrating coupling of photoluminescence from near-surface NV centers to selected devices. This work is motivated not only by the potential for large improvements in quantum efficiency, but also by the inherent scalability advantages of integrated systems.

There has been other work on integrating photonic devices with NV centers [11–14]. However the majority of these devices use the diamond itself as the waveguiding layer. In this work, a thin gallium-phosphide (GaP) waveguiding layer couples to NV centers within ~20 nm of the diamond surface. This hybrid GaP-on-diamond material system [15] yields several advantages. Large-area GaP membranes with uniform thickness can be grown and transferred to diamond. This allows for chip-scale integration of devices with good cross-chip uniformity and high yield, which has proven difficult to achieve by undercutting or thinning the diamond itself [14]. Further, GaP exhibits a second-order optical non-linearity, and thus allows for integration of electro-optic switches based on the linear electro-optic effect [17,18]. Centro-symmetric diamond has no second-order non-linearity. The main disadvantage of using GaP, rather than diamond, as the waveguiding layer is a reduced interaction between NV centers and guided optical modes. However, this effect can be mitigated with resonant devices of sufficient quality factor.

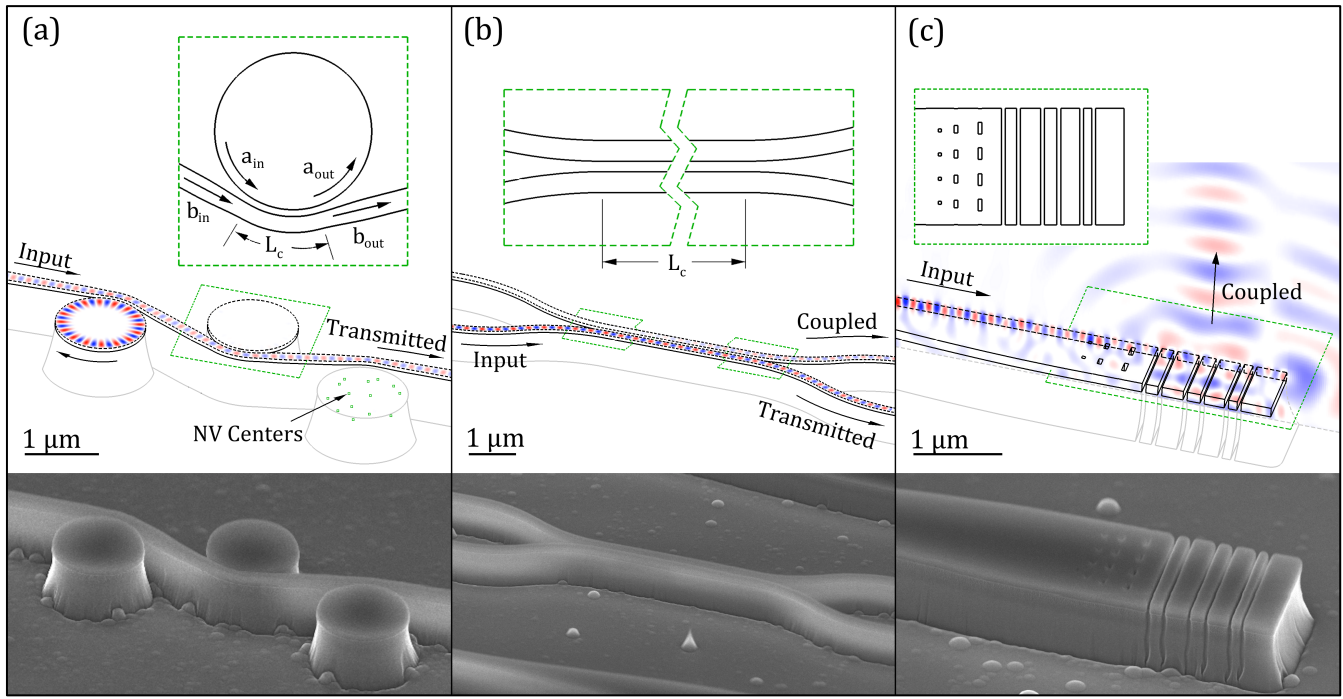

Fig. 1. Schematic views with overlaid FDTD simulations, and SEM images of integrated devices: (a) waveguide-coupled disk resonators; (b) directional coupler; (c) grating coupler (with schematic view halved for clarity).

## 2. FABRICATION

Devices were fabricated using a process similar to that described in our prior work [19]. NV centers were created approximately 15 nm below the surface in a 2 mm x 2 mm commercially available electronic grade diamond chip (Element6) by ion implantation and annealing. The diamond was then cleaned with a solution of $KNO_3$ in fuming $H_2SO_4$, and treated with hexamethyldisilazane (HMDS) vapor.

A 125 nm thick GaP layer on a 300 nm thick $Al_{0.8}Ga_{0.2}P$ sacrificial layer was epitaxially grown on a GaP substrate. A 1.5 mm x 1.5 mm square GaP membrane was released via wet etching in a 1.5% hydrofluoric acid (HF) solution, and transferred to the cleaned and treated diamond surface.

The resulting chip (GaP-on-diamond) was then patterned via electron-beam lithography, using hydrogen silsesquioxane (HSQ) as a negative resist. A $Cl_2/N_2/Ar$ reactive ion etch (RIE) step was performed to etch through the GaP, followed by an $O_2$ RIE step to etch approximately 600 nm into the diamond substrate. The resulting device cross-section consists of the 125 nm GaP layer on top of a 600 nm diamond pedestal, as can be seen in the SEM images shown in Figure 1. An important fabrication difference with regards to our prior work [19] was the addition of $N_2$ into the GaP RIE chemistry to remove a problematic lateral etch bias.

## 3. CHIP-SCALE TRANSMISSION MEASUREMENTS

### A. Measurement Setup

Transmission measurements were taken using a custom-built microscope (see Figure 2 (a)). Excitation was provided by an LED with a peak wavelength of 640 nm, and with a relatively broad emission spectrum extending from approximately 630 nm to 645 nm. Input light was focused onto the sample with a commercially available microscope objective (Nikon, 60x, 0.7 NA). Transmitted light was collected through the same objective and focused through a 100 μm pinhole. The light collected through the pinhole was either coupled into a photodetector or a grating spectrometer, or imaged onto a CCD camera. The input and collection optical paths were separated by a 50:50 beam-splitter.

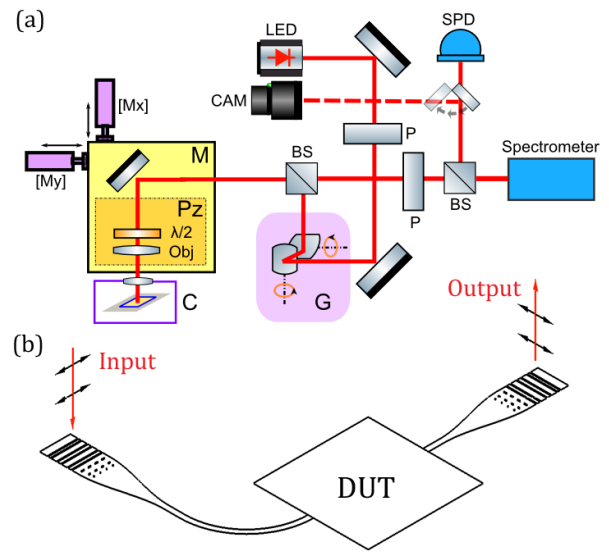

Fig. 2. (a) Schematic of microscope setup, showing galvo mirror system (G), micrometer stage (M), and piezo stage (Pz) used for automated testing. Additional labels: P: polarizer; BS: beam-splitter; SPD: photodetector; C: cryostat sample chamber; λ/2: half-wave plate. (b) Schematic of transmission measurement for an arbitrary device-under-test (DUT).

Light was coupled to and from the chip by means of grating-couplers (see Figure 2(b)) [20]. In order to reduce collection of scattered light from the input grating, input and output gratings were defined at right angles to one another, and the input and collection paths of the microscope were cross-polarized. Input polarization was controlled via a half-wave plate on the objective side of the 50:50 beam-splitter, with all measurements being done for TE-polarized guided modes.

Microscope automation was crucial for obtaining the large amount of transmission data presented in the remainder of this section. Linear actuators and piezoelectric-controlled stages allowed for large-scale and fine movement of the microscope objective over the chip, respectively. Galvo-controlled mirrors in the input beam path allowed scanning of the input beam while collecting from a fixed area. These elements, combined with a grid-based device layout allowed for automated testing of all devices of a certain class in a single testing run, robust to device-to-device variations.

### B. Waveguide-Coupled Disk Resonators

Because the success rate of measurement-based entanglement scales as the square of total quantum efficiency [3], and because demonstrated efficiencies have been low [8,9], there remains a great deal of room for improved entanglement rates through photonic integration. There are two reasons for this. First, NV centers exist in diamond, a high-index material (n = 2.4), making extraction of light into free space difficult. Even using solid-immersion lenses (SILs), the photon collection efficiency increases to only about 5% [21]. Second, only photons emitted at the zero-phonon line (ZPL) wavelength can be used for the implementation of measurement-based quantum protocols. ZPL photons make up approximately 3% of the total emission from an NV center, with the remainder emitting in the phonon sideband (PSB) [8]. One way to overcome both of these effects is the integration of in-plane, resonant photonic devices directly on the surface of the diamond.

In-plane resonant structures can be used to enhance emission into the ZPL through the Purcell effect [22]. Photonic crystal cavities built around NV centers have been demonstrated with Purcell factors as high as $F_p = 69$ [13], corresponding to ~68% of emission into the ZPL. Further, for $F_p \gg 1$, nearly all ZPL photons are emitted into the resonator mode [19]. The Purcell factor of an optical resonator scales linearly with the quality factor, Q, and is inversely proportional to the resonant mode volume. Equally important for the purposes of NV-center based quantum information is that ZPL photon emission be coupled into a useful optical mode. Thus, a second important metric for a resonant device is the coupling efficiency to an output mode. The output coupling efficiency of a coupled disk resonator scales as the square of the field coupling coefficient κ.

**Table 1. Resonator Coupling Geometries**

| Type | Bus Width (nm) | Spacing (nm) | Lengths (rad) |
|---|---|---|---|
| 1 | 160 | 80 | 0.1; 0.2; 0.3 |
| 2 | 140 | 80 | 0.175; 0.35; 0.525 |
| 3 | 130 | 100 | 0.475; 0.95; 1.425 |

Waveguide-coupled disk resonators, designed to provide both a significant Purcell enhancement as well as a high output coupling efficiency, were fabricated in the GaP-on-diamond platform. The resonators consist of whispering-gallery mode disks, with a coupling region in which a bus waveguide is brought into close proximity of the disk (see Figure 1(a)). Three types of coupling region were used, as defined by bus waveguide width and disk-to-bus spacing. For each type of coupler, disks were built in triplets along individual bus waveguides, with the length of the coupling region, $L_c$, varying within each triplet. Disk diameter was also slightly varied (±4 nm) within each triplet in order to distinguish individual disk resonances. An SEM image of a disk triplet is shown in Figure 1(a). A summary of the different coupling geometries can be found in Table 1.

Transmission measurements were taken on the disk resonators, with transmitted light coupled to the spectrometer. Measurement loops were determined to be working by the presence of a measurable output from the output grating coupler. The yield for full measurement loops was 97 out of 109 (89%). Within working measurement loops, waveguide-coupled disks were determined to be working by the presence of an associated resonance dip in the normalized transmission spectrum. The yield for waveguide-coupled disks was 127 out of 219 (58%). An example transmission spectrum for a measurement loop with 3 working disks is shown in Figure 3.

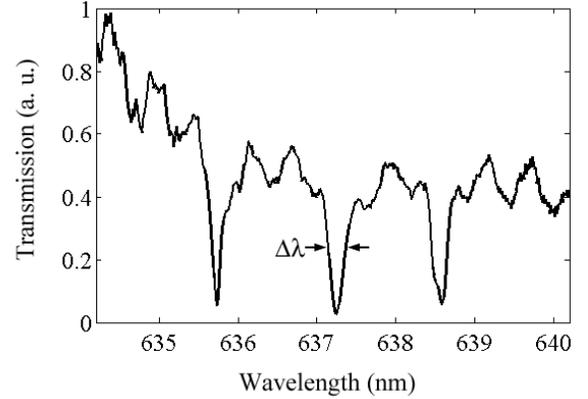

Fig. 3. Measured transmission spectrum for a disk measurement loop with 3 working disks. These devices used Coupler Type 2 (see Table 1).

The quality factor of a coupled disk resonator is a measure of its ability to store optical energy, and is limited both by intrinsic losses and out-coupling to the bus waveguide. Here, we define the quality factor in terms of measured quantities as:

$$Q = \frac{\lambda_{res}}{\Delta\lambda} \quad (1)$$

where $\lambda_{res}$ is the resonant wavelength and $\Delta\lambda$ is the full width at half power of the observed resonance. The extinction ratio, $r_{ext}$, for a coupled disk resonator is a measure of how well coupled it is, and is defined in terms of measured quantities as:

$$r_{ext} = 1 - \frac{T_{res}}{T_0} \quad (2)$$

where $T_{res}$ and $T_0$ are the normalized transmission through the bus waveguide, on and off of resonance respectively. The field coupling coefficient, κ, for a waveguide-coupled disk can be calculated from the measured quality factor and extinction ratio. It is defined in terms of the behavior of the coupling region, for which the output fields are related to the input fields by:

$$\begin{bmatrix} a_{out} \\ b_{out} \end{bmatrix} = \begin{bmatrix} t & i\kappa \\ i\kappa & t \end{bmatrix} \begin{bmatrix} a_{in} \\ b_{in} \end{bmatrix} \quad (3)$$

where $a_{in}, b_{in}, a_{out}$ and $b_{out}$ are the (disk and bus) input and output mode field coefficients, and t is the field transmission coefficient (see blue inset in Figure 1(a)) [23].

Quality factors and extinction ratios were determined by fitting resonant dips in the measured transmission spectra. Figure 4 shows box-and-whisker plots of the measured quality factors and extinction ratios, and calculated coupling coefficients.

For the weakest disk-to-bus coupling geometries (Coupler Type 1), all of the devices were clearly under-coupled. However among devices with stronger coupling geometry (Coupler Types 2 and 3), there were many near-critically-coupled and over-coupled devices. It is also worth noting that while Coupler Types 2 and 3 showed similar coupling coefficients, the average calculated intrinsic quality factor for devices using Coupler Type 3 was significantly higher ($Q_i$ = 11 700), than for those using Coupler Type 2 ($Q_i$ = 8400). This is likely due to the wider disk-to-bus spacing used in Coupler Type 3, and suggests that coupler loss is playing a role in limiting the intrinsic quality factors of the disks.

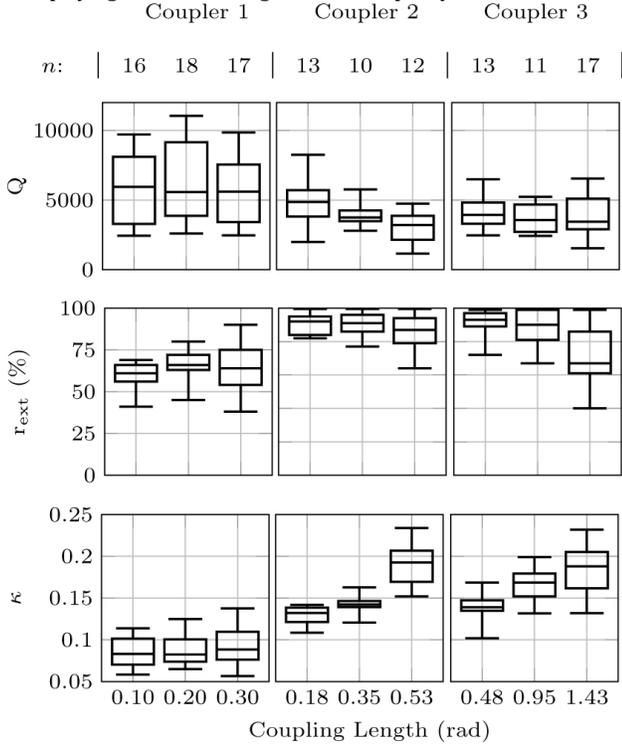

Fig. 4. Box-and-whisker plots showing the distribution in measured quality factor, measured extinction ratio, and calculated coupling coefficient for devices as a function of coupling geometry. Sample size for each coupling geometry is indicated at the top of the corresponding column.

One of several promising devices for ZPL photon collection had a measured quality factor of Q = 5100 and a measured extinction ratio of 67%. The device used Coupler Type 3 with a coupling length of 1.425 radians, and was determined to be over-coupled based on the statistical behavior of similar devices. Given the simulated field distribution for the resonator mode, this would correspond to a Purcell factor of approximately 24 for an ideally oriented NV center at a depth of 20 nm below the diamond surface, or 44% of total photons emitted at the ZPL wavelength. 96% of these would emit into the resonator mode. The disk-to-bus coupling efficiency was calculated to be 79% for this device. The potential total quantum efficiency of ZPL photon collection into the bus waveguide is thus approximately 33%. This is over 200 times higher than what has been demonstrated using free-space optics.

**C. Directional Couplers**

Directional couplers allow low-loss, arbitrary coupling between two waveguide modes, and can be thought of as the integrated optics equivalent of a free-space beam-splitter or beam-sampler [24,25]. They can be used to erase "which-path" information when performing 2-qubit quantum protocols such as those described in [8–10], and are therefore highly desirable in an integrated system for NV center-based MBQC. Further, they can be used as beam samplers with arbitrary coupling ratios throughout an integrated system for testing. In both cases, well-defined coupling ratios and low excess insertion loss are desirable. In particular, a 50:50 coupling ratio is required to obtain a maximally entangled state without sacrificing efficiency when performing spin entanglement between NV centers [3]. However, devices with non-ideal coupling ratios can still be used to obtain a maximally entangled state, but with total quantum efficiency scaling as the lesser of either the coupling or transmission ratio. E.g. devices with either a 40:60 or 60:40 coupling-to-transmission ratio could be used to obtain maximally entangled states, but with total quantum efficiency scaled by a factor of 0.4.

Directional couplers were fabricated in the GaP-on-diamond platform, consisting of narrow ridge waveguide sections spaced closely together (Figure 1(b)). Two types of couplers were fabricated and tested, based on 180 nm ridge waveguides and 160 nm ridge waveguides respectively. Both types had an 80 nm inter-waveguide spacing. For each type, 9 evenly distributed coupling lengths ($L_c$) were fabricated and organized into localized sets of 9.

Transmission measurements were taken on the directional couplers. Transmitted and coupled optical modes were measured at two separate outputs, with output light being coupled into a photodetector. Working measurement loops were defined as those with total output power (coupled + transmitted) at least 25% that of the lowest-loss devices. The yields were 76 out of 103 (74%) and 75 out of 104 (72%) for measurement loops with 180 nm ridge couplers and 160 nm ridge couplers, respectively.

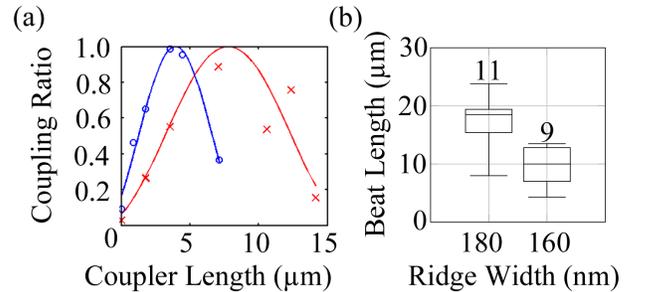

Fig. 5. (a) Measured coupling ratios (markers) and sinusoidal fits (curves) for sets of 180 nm couplers (red, with X markers) and 160 nm couplers (blue, with circular markers). (b) A box-and-whisker plot showing the cross-chip distribution of estimated beat lengths for couplers of each type. The number of sets represented is noted above each distribution.

Transmission and coupling ratios were obtained for each working loop, normalized to the total output. For sets with more than 5 working devices, the coupling ratios were fit to sinusoids in order to obtain a characteristic beat length. The beat length is defined as the distance required for power to couple completely from one mode to the other, and back again. Selected data and associated fits are shown for sets of each geometry in Figure 5(a). The average beat lengths were 17.1 μm and 9.4 μm for the 180-nm ridge geometry and 160-nm ridge geometry, respectively. A box-plot of calculated beat length for both geometries is shown in Figure 5(b). The observed set-to-set variation in beat length suggests a 180 nm ridge-based coupler with a length of

2.2 μm can be expected to fall within the range of 0.4 to 0.6 coupling ratio approximately 60% of the time.

Excess loss in the couplers was determined to be 0.063 ± 0.015 dB/μm and 0.060 ± 0.025 dB/μm, for the 180-nm and 160-nm devices respectively. We can thus put an upper bound of 0.17 dB on the excess insertion loss for 180-nm ridge-based coupler with a length of 2.2 μm.

A low-loss directional coupler with a coupling ratio in the range of 0.4 to 0.6, when combined with a pair of resonators such as the one described at the end of section 3.B, could be used for entanglement generation with a total quantum efficiency of 13% into a single on-chip guided mode. This represents nearly two orders of magnitude improvement in total quantum efficiency, compared with free-space. Based on the measured yields of the individual components and the probability of the directional coupler falling in the desired range, we can expect a yield of approximately 1 in 7 for such a system.

### D. Grating Couplers

Grating couplers [20] are the primary method for coupling free-space light to on-chip guided modes in the platform. They are a crucial component for testing and development, and both high efficiency and cross-chip uniformity are desirable. Higher grating-coupler efficiency translates directly to stronger signals, and therefore shorter required measurement times. This is particularly true in the case of transmission measurements where the transmitted optical power scales as the square of grating coupler efficiency. Uniformity is important, as uncertainty in grating coupler performance can translate into uncertainty in the measurement.

Grating couplers were designed and fabricated on the GaP-on-diamond chip (see Figure 1(c)). Design was performed using a combination of finite-difference time domain (FDTD) and scattering matrix simulations. The designed-for output mode was a 750 nm free-space Gaussian beam propagating normal to the chip surface.

Grating coupler transmission measurements were performed. The measurement loop structures were quite simple: two grating couplers joined by a short section of single-mode waveguide. Output light was coupled into a spectrometer. The yield for measurement loops was 33 out of 42 (79%), where loops were determined to be working if the transmitted power was within 2 standard deviations of the mean. We note that the yield for individual gratings is likely higher than that of the measurement loops since loop failures can be caused by the failure of a single grating or a failure in the bus waveguide.

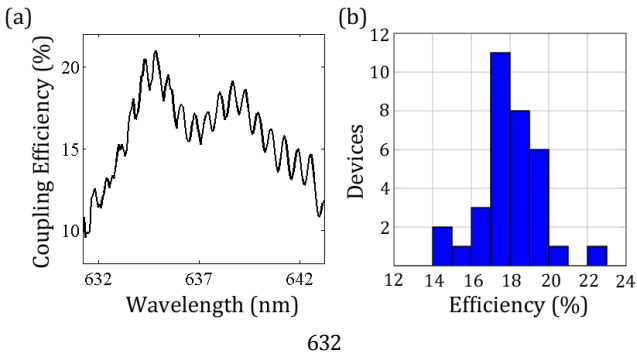

Fig. 6. (a) A typical grating coupler efficiency spectrum. (b) A histogram showing the distribution of grating efficiencies at the ZPL wavelength (637.2 nm).

Figure 6(a) shows a typical efficiency spectrum for a measurement loop, where coupling efficiency is taken as the square root of transmission. Figure 6(b) shows a histogram of coupler efficiencies at the ZPL wavelength of 637.2 nm. The average coupling efficiency for the grating couplers at 637.2 nm was 17.0%, with a cross-chip standard deviation of 1.5%.

The demonstrated cross-chip uniformity will in many cases be good enough to directly compare device performance across the chip, and local calibration can likely be used to improve uncertainty in grating performance when necessary. The measurement efficiencies, while well below unity, are sufficient for quantum device measurements including photon correlation measurement, resonant excitation spectroscopy, and measurement of Purcell enhancements. Ultimately this will be their primary role, as an eventual fully integrated quantum information circuit should make use of on-chip photon detection [26,27].

Nonetheless, when paired with a resonator such as the one described at the end of section 3.B a grating coupler with average coupling efficiency would enable total off-chip quantum efficiency as high as 5.5%. This represents a factor of 36 improvement over free-space collection with SILs.

### 4. COUPLED NV CENTER EMISSION

As an initial demonstration in this platform, coupled NV center emission measurements were performed on selected devices. For these measurements, a diode-pumped solid-state laser emitting at a wavelength of 532 nm was used to excite near-surface NV centers beneath the fabricated devices. All emission data was taken at low temperature (T ≈ 7K).

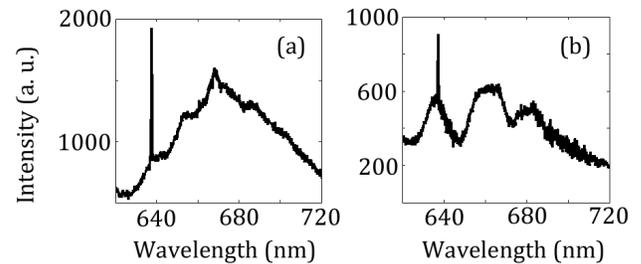

Fig. 7. Spectra taken while exciting NV centers beneath a ridge waveguide, collecting (a) directly from the excitation spot and (b) from an output grating coupler at the end of the waveguide, approximately 10 μm from the excitation spot.

In one setup, the excitation was focused on a section of ridge waveguide. Spectra were collected directly from the excitation spot, as well as from an output grating coupler approximately 10 μm away. As can be seen in Figure 7, both spectra show a clear line at the NV ZPL wavelength, indicating that NV emission is being coupled into the waveguide. Also visible in the top-collected spectrum is the PSB of the NV center. The grating-collected spectrum shows emission in the PSB, but the spectral shape is distorted by the grating coupler. Significantly, the ZPL intensity obtained with grating-collection is about 30% of that obtained with top-collection. Given an average grating coupler efficiency of 17%, this result suggests that the excited NV centers are emitting nearly twice as many photons into the ridge waveguide as into the numerical aperture of the free-space microscope objective.

Similar measurements were performed while optically exciting on a disk resonator. Initially, a photoluminescence image of a triplet of resonators was obtained by scanning the input beam over the devices, and collecting from an output grating coupler. The resulting image shows the whispering-gallery modes around the outside of the resonators (Figure 8(a)), with a particularly bright area indicating one or several well-coupled emitters. Subsequent spectra were taken while exciting on the bright spot, as indicated. Figures 8(b, c) show the spectrum collected directly from the excitation spot, and the spectrum collected from the grating, respectively. The top-collected spectrum shows characteristic NV center emission, as expected. In the grating-collected spectrum, we see only the waveguide-coupled disk resonance. We expect that the light emitted into this mode is from the NV center(s). However only light at the resonant wavelength can couple to the output mode, resulting in the absence of the ZPL. These measurements, while only qualitative in nature, serve to demonstrate that NV center emission does indeed couple to the GaP waveguiding layer in the devices.

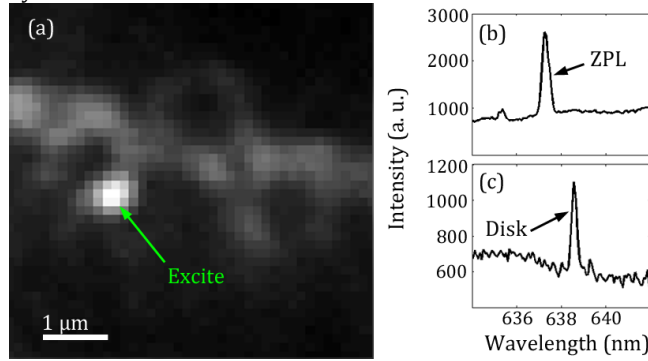

Fig. 8. (a) Grating-coupled photoluminescence image of disk triplet. (b) Photoluminescence spectrum obtained while both exciting and collecting at the spot indicated in (a). (c) Grating-collected photoluminescence spectrum obtained while exciting at the spot indicated in (a).

## 5. DISCUSSION

We have presented chip-scale transmission measurements of three key components of an integrated photonics platform for NV center-based quantum information, as well as proof-of-principle measurements indicating that near-surface NV center emission couples into the devices. This type of large-scale integration and testing is expected to be critical to the development of scalable NV center-based MBQC. Yield and performance of integrated classical optical components will need to be well characterized and optimized in order to facilitate the more difficult-to-engineer coupling of single NV centers to individual resonators.

**Table 2. Expected Yield and Efficiency for Simple 2-Qubit Systems**

| System | T. Q. E. | Yield | T.Q.E. (on-chip) |
|---|---|---|---|
| 2 x (Disk → Grating) | 5.5% | 21% | 33% |
| (2 x Disk) → D.C. → Grating | 2% | 10% | 13% |

While the primary goal of this work was to set a benchmark for large-scale integration in the GaP-on-diamond material platform, the demonstrated device performance suggests there are already areas where the platform should provide large advantages relative to free-space techniques. The measured quality factors of the waveguide-coupled resonators, along with relatively large output coupling, should allow for total quantum efficiency into useful waveguide modes of around 33%. This represents a potential improvement by a factor of more than 200 over what has been demonstrated with free-space optics. Further, the combined yields and device performance of the three demonstrated components are sufficient to allow the integration of elements for a spin entanglement protocol, with significant total quantum efficiency advantages and reasonable yield. Table 2 summarizes the expected total quantum efficiency and full-circuit yield for 2 simple systems. The first system consists of two separate collection channels, each made up of a waveguide-coupled resonator and an output grating, The second system consists of two waveguide-coupled resonators connected to a directional coupler, with a single grating coupler at one of the output ports.

The quantitative device performance and yields in this work are promising for the success of large scale MBQC. The critical next step will be to engineer the quantum properties of the device-integrated NV centers. NV centers created via implantation and annealing have typically exhibited degraded optical properties relative to NV centers incorporated during diamond growth [28]. However we are encouraged by recent significant progress toward improving the properties of near-surface NV centers [29].

We also note that the implementation of measurement-based entanglement generation in the platform will likely require the integration of Stark tuning [30], and RF fields for spin manipulation [8]. In addition, scalable MBQC in the GaP-on-diamond platform will likely require the integration of on-chip single photon detection and electro-optic switching. We expect these additional functionalities to be compatible with the platform, and that a well-characterized suite of photonic devices will aid significantly in their development.

**Funding.** National Science Foundation (NSF) (1343902, 1506473); European Commission(FP7-ICT-2013-613024-GRASP).

**Acknowledgment**. Fabrication was performed at the Washington Nanofabrication Facility (WNF). We thank Richard Bojko and Andrew Lingley in particular for help with the electron-beam lithography system and the RIE system, respectively. We also thank Yaxuan Zhou for helpful FDTD simulations.